\documentclass[aps,prd,reprint,onecolumn,showpacs,preprintnumbers, amsmath,amssymb,floatfix, nofootinbib,superscriptaddress]{revtex4-2}
\usepackage{graphicx}
\usepackage{dcolumn}
\usepackage{bm}
\usepackage{amsmath}

\usepackage[dvipsnames]{xcolor}

\definecolor{persianblue}{rgb}{0.11, 0.22, 0.73}

\usepackage[
	colorlinks=true,
	citecolor=green!50!black,
        linkcolor=persianblue!70!black,
	urlcolor=green!50!black,
	hypertexnames=false]{hyperref}

\usepackage[mathlines]{lineno}
\usepackage{color}
\usepackage{xcolor}

\begin{document}

\title{Revisiting $C$ and $CP$ Violation in $\eta \to \pi^+\pi^-\pi^0$ Decay}
\author{Jun Shi}
\affiliation{Key Laboratory of Atomic and Subatomic Structure and Quantum Control (MOE), Guangdong Basic Research Center of Excellence for Structure and Fundamental Interactions of Matter, Institute of Quantum Matter, South China Normal University, Guangzhou 510006, China}
\affiliation{Guangdong-Hong Kong Joint Laboratory of Quantum Matter, Guangdong Provincial Key Laboratory of Nuclear Science, Southern Nuclear Science Computing Center, South China Normal University, Guangzhou 510006, China }

\author{Jian Liang}
\email{jianliang@scnu.edu.cn}
\affiliation{Key Laboratory of Atomic and Subatomic Structure and Quantum Control (MOE), Guangdong Basic Research Center of Excellence for Structure and Fundamental Interactions of Matter, Institute of Quantum Matter, South China Normal University, Guangzhou 510006, China}
\affiliation{Guangdong-Hong Kong Joint Laboratory of Quantum Matter, Guangdong Provincial Key Laboratory of Nuclear Science, Southern Nuclear Science Computing Center, South China Normal University, Guangzhou 510006, China }

\author{Susan Gardner}
\affiliation{Department of Physics and Astronomy, University of Kentucky, Lexington, KY 40506-0055, USA}

\begin{abstract}
The decay $\eta\to \pi^+\pi^-\pi^0$ is an ideal process 
in which 
to study flavor-conserving $C$ and $CP$ violation beyond the Standard Model.
We deduce the $C$- and $CP$-odd quark operators that contribute to $\eta\to \pi^+\pi^-\pi^0$ 
originating from the mass-dimension 6 Standard Model effective field theory.
The corresponding hadron-level operators that 
generate a 
non-vanishing $I=0$ amplitude at order $p^6$ in the chiral effective theory are presented for the first time, 
in addition to 
the leading order operators
ascribed 
to the $I=2$ final state.
By fitting the KLOE-2 and the most recent BESIII experimental data, we determine the coefficients of the lowest order $I=0$ and $I=2$ amplitudes and estimate the potential new physics energy scale.
We also perform an impact study of the future $\eta\to \pi^+\pi^-\pi^0$ experiments.

\end{abstract}

\maketitle

\section{Introduction}
$C$ and $CP$ violation 
have long been regarded
as 
essential conditions in baryogenesis~\cite{Sakharov:1967dj}, 
that generates the 
observed 
matter-antimatter asymmetry in universe~\cite{Planck:2018vyg}.
However, this baryon asymmetry of universe (BAU) cannot be 
explained within the 
Standard Model (SM), because its electroweak phase transition can not be of first order~\cite{Kajantie:1995kf,Csikor:1998eu} and its mechanism of 
$CP$ violation, through a single
phase in the CKM matrix,  
is far from sufficient~\cite{Farrar:1993sp,Farrar:1993hn,Huet:1994jb,Gavela:1994dt,Elor:2018twp} to explain its size, though exceptions may  
exist~\cite{Elor:2018twp,Nelson:2019fln}.
This motivates the continuing search for new $CP$-violating sources beyond the SM 
(BSM).
Previous BSM $CP$-violation studies 
have focused mainly on flavor-changing processes, such as in $B$, $D$, 
and $K$ meson decays, 
e.g.~\cite{Carter:1980hr,Grossman:1996ke,Laplace:2002ik,Giri:2003ye,Kundu:2005jd,Botella:2005ks,Mohapatra:2019wcm, 
LHCb:2019hro,Dery:2019ysp,Bause:2020obd,Grossman:1997sk, Buchalla:1998ba},
and on electric dipole moments, 
e.g.~\cite{Lamoreaux:2009zz,Abel:2020pzs,ACME:2018yjb,Roussy:2022cmp,Alarcon:2022ero,Dekens:2018bci,Idegawa:2023bkh,DiLuzio:2023ifp},
which are flavor-conserving $P$- and $T$ ($CP$)-violating observables.
But studies 
of 
flavor-conserving $C$- and $CP$-violation are scarce.
Considering the fact that the SM 
$CP$-violating mechanism lies in the flavor-changing weak coupling, the flavor-conserving $C$- and $CP$-violation might be more sensitive to physics BSM. 
An explicit example would be
a charge asymmetry in 
$\eta\to\pi^+\pi^-\pi^0$ decay.

The decay $\eta\to\pi^+\pi^-\pi^0$ is flavor conserving, and its 
parity is conserved due to angular momentum conservation.
This decay can occur via either 
$C$-conserving but isospin-breaking or $C$-violating processes~\cite{Lee:1965zza}.
The latter comes from the interference of the $C$-conserving and the $C$-violating amplitudes, which is proportional to the BSM $C$- and $CP$-violating coefficients~\cite{Gardner:2019nid},
rather than the coefficients squared as in the branching ratio of a pure $CP$-violating process.
Thus 
new physics 
may be more appreciable in
$\eta\to\pi^+\pi^-\pi^0$ decay, 
making 
this channel 
an ideal arena to study flavor-conserving $C$ and $CP$ violation at low energy.

The possibility of $C$ and $CP$ violation 
in 
$\eta\to\pi^+\pi^-\pi^0$ first received attention after the discovery of $CP$ violation through the $K_L\to\pi^+\pi^-$ decay in 1964~\cite{Christenson:1964fg}, to test the validity of the proposed new interaction~\cite{Lee:1965hi,Prentki:1965tt} inducing the $CP$-violating $K_L$ decay.
However, at that time the $C$-conserving amplitude was incorrectly believed to go through a virtual electromagnetic interaction~\cite{Lee:1965zza,Nauenberg:1965cqj,Barrett:1965ia}, which 
was later proved to be
negligibly small~\cite{Sutherland:1966zz,Bell:1968wta,Wilson:1969zs,Baur:1995gc,Ditsche:2008cq}, and the process is dominated by strong interaction~\cite{Gasser:1984pr,Anisovich:1996tx}.
For over five decades since then, 
$C$ and $CP$ violation in $\eta\to\pi^+\pi^-\pi^0$ decay was not further investigated by theorists, with theoretical studies focusing on better descriptions of the final-state interactions within the SM to extract 
the light-quark mass ratio precisely~\cite{Gross:1979ur,Langacker:1978cf,Leutwyler:1996qg,Bijnens:2002qy,Bijnens:2007pr,Colangelo:2009db,Kampf:2011wr,Guo:2015zqa,Albaladejo:2017hhj,Colangelo:2016jmc,Colangelo:2018jxw}.
Very recently, 
theoretical studies of $C$ and $CP$ violation in $\eta\to\pi^+\pi^-\pi^0$ 
decay 
have been made by Gardner and Shi~\cite{Gardner:2019nid} and Akdag, Isken, and Kubis~\cite{Akdag:2021efj},  using different phenomenological frameworks,
in which similar patterns of the $C$-violating amplitudes with $I=0$ and $I=2$ final states are 
obtained.
In searching for the origin of the $CP$-violating mechanism, Refs.~\cite{Shi:2017ffh,Gardner:2024abc} derive all the $C$- and $CP$-odd quark-level operators from the dimension-6 SM 
Effective Theory (SMEFT)~\cite{Grzadkowski:2010es}, while Ref.~\cite{Akdag:2022sbn} lists similar operators based on low energy effective theory (LEFT).
The latter work also matches the operators to hadron-level in chiral perturbation theory (ChPT), and by
combining with the results of Ref.~\cite{Akdag:2021efj}, a naive dimensional analysis (NDA) of the new physics scale has been carried out. 
Here we improve upon this first 
analysis through the use of SMEFT, as
we shall explain. 

Experimentally, the 
charge asymmetry in $\eta\to\pi^+\pi^-\pi^0$ decay 
can 
be identified by observing the 
population asymmetry $A_{LR}$ in the Dalitz plot distribution with respect to the mirror line 
$t=u$ with $t\equiv (p_{\pi^-}+p_{\pi^0})^2$ and $u\equiv (p_{\pi^+}+p_{\pi^0})^2$~\cite{Layter:1972aq}.
Other $C$-asymmetry observables include the quadrant asymmetry $A_Q$ and sextant asymmetry $A_S$, which 
probe 
the final $I=2$ and $I=0$ states~\cite{Lee:1965zza,Nauenberg:1965cqj}, respectively.
Except for three early experiments~\cite{Baltay:1966zz,Gormley:1968zz,Gormley:1970qz} that reported $C$-asymmetry signals,  
with a significance 
of less than $3\sigma$,
no $C$-violation is observed in other experiments~\cite{Larribe:1966zz,Layter:1972aq,Layter:1973ti,Jane:1974mk,KLOE:2008tdy,WASA-at-COSY:2014wpf,BESIII:2015fid}, including the recent high statistics measurements from KLOE-2 collaboration~\cite{KLOE-2:2016zfv} and BESIII~\cite{BESIII:2023edk}.
Future $\eta$ related experiments 
with 
much higher statistics 
from
the 
JLab eta factory (JEF) experiment~\cite{Gan:2009zzd,Gan:2015nyc,Gan:2017kfr,Somov:2024jiy}, REDTOP~\cite{Gatto:2016rae,Gatto:2019dhj,REDTOP:2022slw} collaboration, and the eta factory of the  High-Intensity heavy-ion Accelerator Facility (HIAF) in China~\cite{Chen:2024wad} are on the way.

In this work, we begin with the BSM $C$ and $CP$-odd quark-level operators pertinent 
to $\eta\to\pi^+\pi^-\pi^0$ decays from the dimension 6 SMEFT~\cite{Grzadkowski:2010es}, in which the coefficients are suppressed by the new physics scale as $1/\Lambda^2$.
This is different from the $1/\Lambda^4$ scaling behavior in the LEFT work~\cite{Akdag:2022sbn}, because LEFT cannot resolve the
difference between $\Lambda$ and the mass of the SM weak
gauge bosons. 
Next, we develop 
the hadron-level operators at order $p^2$ in ChPT accounting for the $I=2$ final state, 
as well as the lowest-order operators generating a non-vanishing $I=0$ amplitude at order $p^6$. Although 
examples of the former
are presented in Ref.~\cite{Akdag:2022sbn}, 
the latter are presented here for the first time.
With the $I=0$ and $I=2$ ChPT amplitudes, we fit the KLOE-2 data~\cite{KLOE-2:2016zfv} together with the most recent BESIII data~\cite{BESIII:2023edk} directly to obtain the coefficients of the order $p^2$ and $p^6$ amplitudes, which are used to estimate the potential new physics scale.
Finally, an impact study of future $\eta\to\pi^+\pi^-\pi^0$ experiments is carried out, which provides some guidance to the upcoming $\eta$ factory experiments~\cite{Gan:2009zzd,Gan:2015nyc,Gan:2017kfr,Somov:2024jiy,Gatto:2016rae,Gatto:2019dhj,REDTOP:2022slw, Chen:2024wad}.
According to our study, it 
should be 
possible to observe $C$- and $CP$-violation 
in 
$\eta\to\pi^+\pi^-\pi^0$ decay in the near future.

This paper is organized as follows. In Sec.~\ref{sec:theoretical_form}, we introduce the quark-level operators and the matching to ChPT operators contributing to $C$ and $CP$ violation of $\eta\to\pi^+\pi^-\pi^0$.
In Sec.~\ref{sec:method}, we explain the fitting procedure to determine the coefficients of $C$-violating amplitudes with specific final state isospin.
Then we carry out NDA of the new physics scale from these coefficients and also an impact study for future experiments in Sec.~\ref{sec:new_physics_scale}.
A brief summary is provided 
in Sec.~\ref{sec:summary}. 

\section{$C$- and $CP$-odd operators}\label{sec:theoretical_form}

In this section, we first briefly recall the derivation of the $C$- and $CP$-odd flavor-conserving quark-level operators from SMEFT, then show their matching to ChPT operators,
from which the lowest-order $I=2$ and $I=0$ $C$-violating amplitudes of $\eta\to\pi^+\pi^-\pi^0$ are obtained. 
 
The SMEFT Lagrangians reads
\begin{equation}
\mathcal{L}=\mathcal{L}_{SM}^{(4)}+\sum\limits _{i}\frac{C_{i}}{\Lambda^{D-4}}\mathcal{O}^{D},
\end{equation}
where the operators $\mathcal{O}^{D}$ have mass dimension $D>4$ with a suppression by the new physics scale $\Lambda$ and $C_i$ are the corresponding dimensionless Wilson coefficients.
The SMEFT shares the same gauge symmetries and building blocks as the SM, which are the $SU(2)_L$ left-handed doublets $q_{Lp}=\left(u_{Lp}, d_{Lp}\right)^T$ and $l_{Lp}=\left(\nu_{Lp}, e_{Lp}\right)^T$, the right-handed singlets $u_{Rp}$, $d_{Rp}$, and $e_{Rp}$ with $p$ denoting the generation, the Higgs field $\varphi$, the $SU(3)_C$ gauge field $G_\mu$ and the $SU(2)_L\times U(1)_Y$ gauge fields $W^I_\mu$ and $B_\mu$.
In the particular case of $\eta\to\pi^+\pi^-\pi^0$,
the dimension 5 operator~\cite{Weinberg:1979sa}
does not appear 
since it breaks lepton number
and contains only Higgs and lepton fields.
Thus we need to start from dimension 6 operators.
In Refs.~\cite{Shi:2017ffh,Gardner:2024abc} we have already investigated all the dimension 6 SMEFT operators obtained in Ref.~\cite{Grzadkowski:2010es} and derived the $P$- and $CP$-odd as well as the $C$- and $CP$-odd quark-level operators at the scale just below the weak gauge boson mass.
Here we present the 
essential 
procedure for the readers' convenience and list the lowest-dimensional $C$- and $CP$-odd operators that contribute to $\eta\to\pi^+\pi^-\pi^0$. 

The dimension 6 SMEFT operators~\cite{Grzadkowski:2010es} most pertinent to flavor-conserving $C$ and $CP$ violation~\cite{Shi:2017ffh,Gardner:2024abc} 
can be expressed as the $T$-odd combination
\begin{eqnarray}
{{\Omega}}=\frac{i}{\Lambda^2}&&\left\{\bar{q}_{Lp}\sigma^{\mu\nu}\left[{\mathrm {Im}}(C_{quB\varphi}^{pr})B_{\mu\nu}+{\mathrm {Im}}(C_{quW\varphi}^{pr})\tau^{I}W_{\mu\nu}^{I}\right]\tilde{\varphi}u_{Rr}\right.\nonumber\\
&&\left. + \bar{q}_{Lp}\sigma^{\mu\nu}\left[{\mathrm {Im}}(C_{qdB\varphi}^{pr})B_{\mu\nu}+{\mathrm {Im}}(C_{qdW\varphi}^{pr})\tau_{I}W_{\mu\nu}^{I}\right]\varphi d_{Rr}- {\rm H.c.}\right\} ,\label{eq:O_SMEFT}
\end{eqnarray}
where $\tau_I$ are Pauli matrices, $\tilde{\varphi}=i\tau_2\varphi^\ast$, $C^{pr}_{quB\varphi}$, $C^{pr}_{qdB\varphi}$, $C^{pr}_{quW\varphi}$ and $C^{pr}_{qdW\varphi}$ are the related Wilson coefficients with subscripts indicating the operator constituents and superscripts indicating the 
generations 
of the quark fields. 
We have omitted terms with two weak gauge fields since they will prove to be of higher mass dimension.
After the electroweak symmetry is spontaneously broken, so that 
the Higgs field acquires its vacuum-expectation-value (VEV), we rotate $B_{\mu}$ and $W^I_{\mu}$ to the physical fields $W^\pm_{\mu}$, $Z_\mu$,  and $A_\mu$
and obtain
\begin{eqnarray}
   {{\Omega}}&\sim & \frac{\sqrt{2} v i}{\Lambda^2} \left[ {\mathrm {Im}}(C_{q u Z \varphi}^{p p}) 
  \bar{u}_{L p} \sigma^{\mu \nu} u_{R p} 
  \partial_{\mu} Z_{\nu} + {\mathrm {Im}}(C_{q d Z
  \varphi}^{p p}) 
  \bar{d}_{L p} \sigma^{\mu \nu} d_{R p} 
   \partial_{\mu} 
  Z_{\nu}\right.\nonumber\\
  &+&\left. \sqrt{2} {\rm Im}(C_{q u W \varphi}^{p r})\bar{d}_{L p} \sigma^{\mu \nu} u_{R r}
  \partial_{\mu} W^-_{\nu} + \sqrt{2} {\rm Im}(C_{q d W \varphi}^{p r}) \bar{u}_{L p}
  \sigma^{\mu \nu} d_{R r} \partial_{\mu} W^+_{\nu} \right] + {\rm H.c.},\label{eq:qqW}
\end{eqnarray}
where $C_{quZ\varphi}^{pr} = [c_{w}C_{quW\varphi}^{pr}-s_{w}C_{quB\varphi}^{pr}]$,
$C_{qdZ\varphi}^{pr} = -[c_{w}C_{qdW\varphi}^{pr}+s_{w}C_{qdB\varphi}^{pr}]$,
$s_w\equiv \sin{\theta_W}$, $c_w\equiv \cos{\theta_W}$, and $\theta_W$ is the Weinberg angle.
We also omit all the terms with $A_\mu$ because we suppose quark operators
should dominate BSM effects in 
$\eta\to\pi^+\pi^-\pi^0$.
The operators of Eq.~(\ref{eq:qqW}) are $CP$-odd, but they
do not have definite $P$ or $C$ transformation properties since the weak gauge bosons couple to both vector and axial-vector quark bilinears.
We then integrate out $W^\pm$ and $Z$ when the energy is just below their mass and make the following replacements
\begin{eqnarray}
W_{\mu}^{+}  \rightarrow &&\frac{g}{\sqrt{2}M_{W}^{2}}\bar{d}_{Lx}\gamma_{\mu}V_{px}^{\ast}u_{Lp},\label{WZtoquarks}\\
W_{\mu}^{-}  \rightarrow && \frac{g}{\sqrt{2}M_{W}^{2}}\bar{u}_{Lp}\gamma_{\mu}V_{px}d_{Lx},\\
Z_{\mu}  \rightarrow && \frac{g_Z}
{M_{Z}^{2}}\left[ \bar{u}_{Lp}\gamma_{\mu}(\frac{1}{2}-\frac{2}{3}s_{w}^{2})u_{Lp}+\bar{d}_{Lp}\gamma_{\mu}(-\frac{1}{2}+\frac{1}{3}s_{w}^{2})d_{Lp}\right.\nonumber\\
&&\left.+\bar{u}_{Rp}\gamma_{\mu}(-\frac{2}{3}s_{w}^{2})u_{Rp}+\bar{d}_{Rp}\gamma_{\mu}(\frac{1}{3}s_{w}^{2})d_{Rp}\right],\label{eq:WZtoqbarq}
\end{eqnarray}
where $g_Z=g/\cos{\theta_W}$, $V_{px}$ are the Cabibbo-Kobayashi-Maskawa (CKM) matrix elements, and we sum over the generation indices, omitting the $t$ quark.
Finally, we pick out the lowest mass-dimensional flavor-conserving $C$- and $CP$-odd quark-level operators as
\begin{eqnarray}
  {{\Omega}}^{\not C P} & \sim & \frac{i\sqrt{2}  v}{\Lambda^2} \left\{ \frac{g_Z}{4
  M_z^2} (\text{Im} (C_{q u Z \varphi}^{p p}) \bar{u}_p \sigma^{\mu \nu}\gamma_5 u_p +
  \text{Im} (C_{q d Z \varphi}^{p p}) \bar{d}_p \sigma^{\mu \nu}\gamma_5 d_p)
  \partial_{\mu} \left(\bar{d}_r \gamma_{\nu} \gamma_5 d_r - \bar{u}_r \gamma_{\nu} \gamma_5 u_r
  \right) \right.\nonumber\\
  &+&\left. \frac{g}{4 M_W^2} \text{Im} (C_{q u W
  \varphi}^{p r} - C_{q d W \varphi}^{r p}) \left[ \bar{d}_p \sigma^{\mu \nu}
  u_r \partial_{\mu} (\bar{u}_r \gamma_{\mu} V_{r p} d_p) - \bar{u}_r
  \sigma^{\mu \nu} d_p \partial_{\mu} (\bar{d}_p \gamma_{\mu} V_{r p}^{\ast}
  u_r) \right]\right.\nonumber\\
  &-&\left. \frac{g}{4 M_W^2} \text{Im} (C_{q u W \varphi}^{p r} + C_{q d
  W \varphi}^{r p}) \left[ \bar{d}_p \sigma^{\mu \nu} \gamma_5 u_r
  \partial_{\mu} (\bar{u}_r \gamma_{\mu} \gamma_5 V_{r p} d_p) + \bar{u}_r
  \sigma^{\mu \nu} \gamma_5 d_p \partial_{\mu} (\bar{d}_p \gamma_{\mu}
  \gamma_5 V_{r p}^{\ast} u_r) \right] \right\}.\label{eq:C_odd_lowest_D_1}
\end{eqnarray}
Given our interest in $\eta \to \pi^+ \pi^- \pi^0$, we let $p=r=1$ and then, 
according to their flavor structure, we write down the operators in three kinds
\begin{eqnarray}
    &&i\bar{\psi}_i\sigma^{\mu \nu}\gamma_5\psi_i\partial_\mu(\bar{\psi}_j\gamma_{\nu} \gamma_5\psi_j),\\
    &&i\bar{\psi}_i\sigma^{\mu \nu}\gamma_5\psi_j\partial_\mu(\bar{\psi}_j\gamma_{\nu} \gamma_5\psi_i) +i\bar{\psi}_j\sigma^{\mu \nu}\gamma_5\psi_i\partial_\mu(\bar{\psi}_i\gamma_{\nu} \gamma_5\psi_j),~~~~(i\neq j)\\
    &&i\bar{\psi}_i\sigma^{\mu \nu}\psi_j\partial_\mu(\bar{\psi}_j\gamma_{\nu}\psi_i) -i\bar{\psi}_j\sigma^{\mu \nu}\psi_i\partial_\mu(\bar{\psi}_i\gamma_{\nu}\psi_j),~~~~(i\neq j)
\end{eqnarray}
where $\psi_i$ denotes the quark field with flavor $i$ and 
$V_{ud} \approx 1$.
Using the equations of motion, integration by parts, and Fierz identities~\cite{Liao:2020zyx}, 
it can be proved that these three operators can be converted into one single form $\bar{\psi}_i\overset\leftrightarrow{\partial}_\mu\gamma_5\psi_i\psi_j\gamma^\mu\gamma_5 \psi_j$, where we employ the Hermitian derivative 
$i \bar \psi \overset\leftrightarrow{\partial}_\mu \psi \equiv 
\bar \psi i (\partial_\mu \psi) - i (\partial_\mu \bar\psi)\psi$. 
Thus we rewrite Eq.~(7) for use in first-generation processes as 
\begin{equation}
 {{\Omega}}^{\not C P}  \sim \frac{vg}{M_W^2}\frac{1}{\Lambda^2}c_{ij}\bar{\psi}_i\overset\leftrightarrow{\partial}_\mu\gamma_5\psi_i\psi_j\gamma^\mu\gamma_5 \psi_j,\label{eq:quark}
\end{equation}
where the Wilson coefficients $c_{ij}$ are real with flavor indices and we use the coefficient $vg/(M_W^2\Lambda^2)c_{ij}$ since 
$g_Z\approx g$ and $M_W\approx M_Z$.
The structure of this 
quark-level operator, 
which is dimension 7 in itself, has also been reported  in the 
early literature~\cite{Khriplovich:1990ef,Conti:1992xn,Engel:1995vv,Ramsey-Musolf:1999cub,Kurylov:2000ub} and is adopted in Ref.~\cite{Akdag:2022sbn}, but only with SMEFT 
can the new physics scale dependence be properly handled.
Please note that in Ref.~\cite{Akdag:2022sbn}, they start from LEFT and trace their operators back to dimension-8 SMEFT, which 
has a different ($1/\Lambda^4$) dependence on the BSM scale.
 
Now we show the matching of the quark-level operators in Eq.~(\ref{eq:quark}) to ChPT ones; we neglect all QCD evolution effects in evolving
the quark-level operators 
to the chiral matching scale 
$\Lambda_{\chi}$. 
Following Refs.~\cite{Gasser:1983yg,Gasser:1984gg,Graesser:2016bpz,Liao:2019gex,Akdag:2022sbn}, the $C$- and $CP$-odd ChPT operators are constructed by rewriting the quark-level operators in chiral irreducible representations and replacing the quark field<s with chiral building blocks coupled to 
the spurions 
$\lambda^{(\dagger)}_i$ and $\lambda_{L/R,j}$ 
with flavor indices $i,j\in {u,d,s}$.
After carefully examining all the possible forms, we find the following three operators
with nonzero contribution to $\eta\to\pi^+\pi^-\pi^0$ at order $p^2$ 
and at leading order in the number of colors $N_c$:
\begin{eqnarray}
    {\mathcal{L}}^{\not{C}P}_{p^2} &=& \frac{ivg}{ M_W^2}\frac{1}{\Lambda^2}c_{ij} \left[{g}_1\langle(\lambda_{i} D^2 \bar{U}^{\dagger} \bar{U} \lambda_{L,j} \bar{U}^{\dagger}
   + \lambda^{\dagger}_{i} D^2 \bar{U} \bar{U}^{\dagger} \lambda_{R,j} \bar{U}) - {\rm H.c.} \rangle\right.\nonumber\\
   && + \left.{g}_2\langle(\lambda_{i} D_\mu \bar{U}^{\dagger} D^\mu\bar{U} \lambda_{L,j} \bar{U}^{\dagger}
   + \lambda^{\dagger}_{i} D_\mu \bar{U} D^\mu\bar{U}^{\dagger} \lambda_{R,j} \bar{U}) - {\rm H.c.} \rangle \right. \nonumber\\
   && + \left. {g}_3\langle(\lambda_{i}  \bar{U}^{\dagger} D^2\bar{U}\lambda_{L,j} \bar{U}^{\dagger}
   + \lambda^{\dagger}_{i} \bar{U} D^2\bar{U}^{\dagger}\lambda_{R,j} \bar{U}) - {\rm H.c.} \rangle \right].\label{eq:p2_ChPT}
\end{eqnarray} 
Here the large $N_c$ extension~\cite{Rosenzweig:1979ay,DiVecchia:1980yfw,Witten:1980sp,Kawarabayashi:1980dp,Leutwyler:1997yr,Herrera-Siklody:1996tqr,Kaiser:2000gs,Bickert:2016fgy} 
has been 
applied to include $\eta^\prime$ and  $\bar{U}=\text{exp}\left({\bar{\Phi}}/{F_0}\right)$
with
\begin{equation}
\bar{\Phi}= \begin{pmatrix}
        \frac{1}{\sqrt{3}}\eta^\prime+\sqrt{\frac{2}{3}}\eta+\pi^0 & \sqrt{2}\pi^+ & \sqrt{2}K^+\\
        \sqrt{2}\pi^- & \frac{1}{\sqrt{3}}\eta^\prime+\sqrt{\frac{2}{3}}\eta - \pi^0 & \sqrt{2}K^0\\
        \sqrt{2}K^- &\sqrt{2}\bar{K}^0 & \frac{2}{\sqrt{3}}\eta^\prime-\sqrt{\frac{2}{3}}\eta \end{pmatrix} \,,
\end{equation}
which is useful for future use.
The effective coefficients $g_i$ are of mass-dimension 5.
We note a similar result can be found in 
Ref.~\cite{Akdag:2022sbn}, though we differ in the contribution of a $g_3$ term.
With Eq.~(\ref{eq:p2_ChPT}) in hand, the spurions 
can be set to their physical values, $\lambda^{(\dagger)}_u,\lambda_{L/R,u}  = \text{diag}(1,0,0)$, $\lambda^{(\dagger)}_d,\lambda_{L/R,d}  = \text{diag}(0,1,0)$, and $\lambda^{(\dagger)}_s,\lambda_{L/R,s}  = \text{diag}(0,0,1)$.
Expanding $\bar{U}^{(\dagger)}$ to $\bar{\Phi}^4$, these three operators together produce the following $C$- and $CP$-odd $\eta\to\pi^+\pi^-\pi^0$ interaction
\begin{equation}
  \frac{ivg}{ M_W^2}\frac{1}{\Lambda^2 F_0^4} 2
   \mathcal{N}_{p^2} \partial^\mu \pi^0(\pi^+\partial_\mu \pi^- - \pi^-\partial_\mu \pi^+)\eta,\label{eq:p2_eta}
\end{equation}
where
\begin{equation}
    \mathcal{N}_{p^2} = 4\sqrt{\frac{2}{3}}
   (c_{u u} - c_{ud} - c_{d u} + c_{d d})(-{g}_1 + {g}_2 - {g}_3 ).\label{eq:Neta_p2}
\end{equation}
The corresponding amplitude is
\begin{equation}
   \mathcal{M}^{\not{C}}_{p^2}(s,t,u) = i \frac{vg}{ M_W^2}\frac{1}{\Lambda^2F_0^4} \mathcal{N}_{p^2} (t - u) \equiv i \alpha (t - u),\label{eq:amp_p2}
\end{equation}
where $s = (p_{\pi^+}+p_{\pi^-})^2$ and $\alpha$ is of mass-dimension $-2$.
Generally the ChPT amplitude does not map to a certain final state with specific isospin, but we can decompose it into an $I=0$ amplitude and $I=2$ amplitude according to their isospin structures~\cite{Gaspero:2008rs,Akdag:2022sbn} as
\begin{eqnarray}
    \mathcal{M}^{\not C}_{I=0} &=& \frac{1}{\sqrt{6}}\left[-\mathcal{M}^{\not C}(t,s,u) + \mathcal{M}^{\not C}(s,t,u)-\mathcal{M}^{\not C}(u,t,s)\right],\\
    \mathcal{M}^{\not C}_{I=2} &=& -\frac{1}{2\sqrt{3}}\left[\mathcal{M}^{\not C}(t,s,u) + 2\mathcal{M}^{\not C}(s,t,u)+\mathcal{M}^{\not C}(u,t,s)\right].
\end{eqnarray}
Thus we see that 
$\mathcal{M}^{\not C}_{I=0,p^2}=0$ and $\mathcal{M}^{\not C}_{I=2,p^2}=-\sqrt{3}/2\mathcal{M}^{\not{C}}_{p^2}$.
This demonstrates that the order $p^2$ amplitude only contributes to the $I=2$ final state.
Actually, it is well known that the lowest-order nonzero $I=0$ amplitude first appears 
at order $p^6$~\cite{Prentki:1965tt}.
In order to better understand the $I=0$ case,
we also investigate the order $p^6$ ChPT operators that contribute to the 
$I=0$ final state.
Among the numerous order $p^6$ operators, we find that all operators with three $\bar{U}$ yield a vanishing $I=0$ amplitude, whereas 
the following operators with five $\bar{U}$ have a nonzero contribution: 
\begin{eqnarray}
  \mathcal{{L}}^{\not C P}_{p^6}&=&\frac{ivg}{ M_W^2}\frac{1}{\Lambda^2}c_{ij} \left[{f}_1 \langle \lambda
  \partial_{\mu} \partial_{\nu} \partial_{\rho} \bar{U}^{\dagger} U \lambda_L
  \partial^{\mu} \partial^{\nu} \bar{U}^{\dagger} U \partial^{\rho}
  \bar{U}^{\dagger} + \lambda^{\dagger} \partial_{\mu} \partial_{\nu}
  \partial_{\rho} \bar{U} U^{\dagger} \lambda_R \partial^{\mu} \partial^{\nu}
  \bar{U} U^{\dagger} \partial^{\rho} \bar{U} - {\rm H.c.} \rangle \right. \nonumber \\
 &  & \left.+ {f}_2 \langle \lambda \partial_{\mu} \partial_{\nu}
  \partial_{\rho} \bar{U}^{\dagger} U \partial^{\mu} \partial^{\nu}
  \bar{U}^{\dagger} \lambda_R U \partial^{\rho} \bar{U}^{\dagger} +
  \lambda^{\dagger} \partial_{\mu} \partial_{\nu} \partial_{\rho} \bar{U}
  U^{\dagger} \partial^{\mu} \partial^{\nu} \bar{U} \lambda_L U^{\dagger}
  \partial^{\rho} \bar{U} - {\rm H.c.} \rangle  \right.  \nonumber \\
  &  &\left.+ {f}_3 \langle \lambda \partial_{\mu} \partial_{\nu}
  \partial_{\rho} \bar{U}^{\dagger} \partial^{\mu} \partial^{\nu} U
  \bar{U}^{\dagger} \partial^{\rho} U \lambda_L \bar{U}^{\dagger} +
  \lambda^{\dagger} \partial_{\mu} \partial_{\nu} \partial_{\rho} \bar{U}
  \partial^{\mu} \partial^{\nu} U^{\dagger} \bar{U} \partial^{\rho}
  U^{\dagger} \lambda_R \bar{U} - {\rm H.c.} \rangle  \right.   \nonumber \\
  &  & +\left. {f}_4 \langle \lambda \partial_{\mu} \partial_{\nu}
  \partial_{\rho} \bar{U}^{\dagger} \partial^{\mu} \partial^{\nu} U
  \bar{U}^{\dagger} \lambda_R \partial^{\rho} U \bar{U}^{\dagger} +
  \lambda^{\dagger} \partial_{\mu} \partial_{\nu} \partial_{\rho} \bar{U}
  \partial^{\mu} \partial^{\nu} U^{\dagger} \bar{U} \lambda_L \partial^{\rho}
  U^{\dagger} \bar{U} - {\rm H.c.} \rangle  \right.  \nonumber \\
 &  &\left. + {f}_5 \langle \lambda \partial_{\mu} \partial_{\nu}
  \partial_{\rho} \bar{U}^{\dagger} U \lambda_L \partial^{\mu} \partial^{\nu}
  \bar{U}^{\dagger} \partial^{\rho} U \bar{U}^{\dagger} + \lambda^{\dagger}
  \partial_{\mu} \partial_{\nu} \partial_{\rho} \bar{U} U^{\dagger} \lambda_R
  \partial^{\mu} \partial^{\nu} \bar{U} \partial^{\rho} U^{\dagger} \bar{U} -
  {\rm H.c.} \rangle  \right.  \nonumber \\
  &  & \left.+ {f}_6 \langle \lambda \partial_{\mu} \partial_{\nu}
  \partial_{\rho} \bar{U}^{\dagger} U \partial^{\mu} \partial^{\nu}
  \bar{U}^{\dagger} \lambda_R \partial^{\rho} U \bar{U}^{\dagger} +
  \lambda^{\dagger} \partial_{\mu} \partial_{\nu} \partial_{\rho} \bar{U}
  U^{\dagger} \partial^{\mu} \partial^{\nu} \bar{U} \lambda_L \partial^{\rho}
  U^{\dagger} \bar{U} - {\rm H.c.} \rangle  \right. \nonumber \\
  &  & \left.+ {f}_7 \langle \lambda \partial_{\mu} \partial_{\nu}
  \partial_{\rho} \bar{U}^{\dagger} U \partial^{\mu} \partial^{\nu}
  \bar{U}^{\dagger} \partial^{\rho} U \lambda_L \bar{U}^{\dagger} +
  \lambda^{\dagger} \partial_{\mu} \partial_{\nu} \partial_{\rho} \bar{U}
  U^{\dagger} \partial^{\mu} \partial^{\nu} \bar{U} \partial^{\rho}
  U^{\dagger} \lambda_R \bar{U} - {\rm H.c.}\rangle\right],\label{eq:ChPT_p6}
\end{eqnarray}
where $f_i$ are of mass-dimension 1.
 Note that these $p^6$ operators contribute both to $I=0$ and $I=2$ final state after expanding the $\bar{U}$ field.
 The corresponding interaction vertex relevant to $I=0$ final state is
\begin{equation}
    \frac{ivg}{ M_W^2} \frac{1}{\Lambda^2F_0^4} 8
   \mathcal{N}_{p^6} \epsilon_{I J K}
   (\partial_{\mu} \partial_{\nu} \partial_{\rho} \pi^I) (\partial^{\mu}
   \partial^{\nu} \pi^J) (\partial^{\rho} \pi^K) \eta,\label{eq:p6_eta_I0}
\end{equation}
where $I,~J,~K=+,~-,~0$ and
\begin{equation}
    \mathcal{N}_{p^6} = \sqrt{\frac{2}{3}}
   (c_{u u} - c_{ud} - c_{d u} + c_{d d})
   ({f}_1 + {f}_2 + {f}_3 - {f}_4 -
   {f}_5 - {f}_6 + {f}_7).\label{eq:Neta_I0}
\end{equation}
The structure of Eq.~(\ref{eq:p6_eta_I0}) 
is in fact well known, e.g. Ref.~\cite{Prentki:1965tt}; however, it is derived here from SMEFT and ChPT for the first time.
The resultant order $p^6$ $I=0$ amplitude is expressed as
\begin{equation}
    \mathcal{M}^{\not{C}P}_{p^6} = i \frac{vg}{ M_W^2}\frac{1}{\Lambda^2F_0^4} \mathcal{N}_{p^6} (s - t) (u - s) (t - u) \equiv i \beta (s - t)
   (u - s) (t - u),\label{eq:amp_p6}
\end{equation}
where $\beta$ is of mass-dimension $-6$.

In Eqs.~(\ref{eq:amp_p2}) and (\ref{eq:amp_p6}),
the effective coefficients $\alpha$ and $\beta$ depend on the new physics scale $\Lambda$.
Thus determining $\alpha$ and $\beta$ from experimental observables helps to estimate the potential new physics scale.
The total amplitude square that is directly related to the experimental measurements can be written as
\begin{equation}
    |\mathcal{M}(s,t,u)|^2=|\mathcal{M}^C(s,t,u)|^2 + 2\text{Re}\left[\mathcal{M}^C(s,t,u)\cdot\mathcal{M}^{\not C}(s,t,u)^\ast \right]+ \mathcal{O}(\alpha^2, \beta^2),\label{eq:ampsq}
\end{equation}
where $\mathcal{M}^C$ is the $I=1$ $C$-conserving amplitude, and $\mathcal{M}^{\not C}$ indicates the $C$-violating amplitudes.
The second term is the interference of the $C$-conserving and $C$-violating amplitudes which contributes to $C$- and $CP$-violation of $\eta\to\pi^+\pi^-\pi^0$ decay.

\section{Determining $\alpha$ and $\beta$ from Experiments}\label{sec:method}
In this section, we describe our procedures to determine $\alpha$ and $\beta$ from experimental observables.
The experimental Dalitz plot distribution of $\eta\to\pi^+\pi^-\pi^0$ is usually described by variables $X$ and $Y$ defined as
\begin{eqnarray}
X &\equiv& \sqrt{3} \frac{T_{\pi^+} - T_{\pi^-}}{Q_{\eta}} =\frac{\sqrt{3}}{2 m_{\eta} Q_{\eta}} (u - t),\\
Y &\equiv& \frac{3 T_{\pi^0}}{Q_{\eta}} - 1 = \frac{3}{2 m_{\eta} Q_{\eta}} [(m_{\eta} - m_{\pi^0})^2 - s] - 1,\label{eq:XY_stu}
\end{eqnarray}
where $ Q_{\eta} = T_{\pi^+} + T_{\pi^-} + T_{\pi^0} = m_{\eta} - 2 _m{\pi^+} - m_{\pi^0} $, and $T_{\pi^i}$ is the $\pi^i$'s kinetic energy in the $\eta$ rest frame.
Since $X,~Y\in(0,1)$, the amplitude squared can be expanded in $X$ and $Y$ as
\begin{equation}
    |\mathcal{M}(s,t,u)|^2 = N_0(1 + aY + bY^2 + cX + dX^2 + eXY + fY^3 + gX^2Y + hXY^2 + lX^3 + \cdots),\label{eq:ampsq_XY}
\end{equation}
where $N_0$ is a normalization factor and $a,~b,~c,~\cdots$ are called Dalitz plot parameters.
Since the $C$ transformation on the amplitude switches $t$ with $u$, which is equivalent to switching $X$ with $-X$, the nonzero coefficients of terms with $X$ in odd power, i.e., $c,~e,~h$ and $l$ in Eq.~(\ref{eq:ampsq_XY}), would imply $C$ and $CP$ violation 
in this decay.
Ref.~\cite{Gardner:2019nid} constructs the $I=0$ and $I=2$ amplitudes by reassembling the $C$-conserving $I=1$ amplitude from NLO ChPT and fitting the mock data generated using the $C$-violating parameters $c,~e,~h$ and $l$ reported by KLOE-2~\cite{KLOE-2:2016zfv} to determine the decay pattern.
In comparison, the 
work of
Ref.~\cite{Akdag:2021efj} uses dispersion theory to form the amplitudes and fits the whole Dalitz plot distribution of $\eta\to\pi^+\pi^-\pi^0$.
In this work, we construct the $C$-violating amplitudes from SMEFT and ChPT without any phenomenological input, and we adopt two procedures to determine $\alpha$ and $\beta$ from the $\eta\to \pi^+\pi^-\pi^0$ data. 
One is to fit the left-right asymmetric distribution directly, and the other is to use the integrated asymmetries $A_Q$ and $A_S$.
We first 
explain the fitting procedure 
in what follows.

KLOE-2~\cite{KLOE-2:2016zfv} and BESIII~\cite{BESIII:2023edk} provide the binned Dalitz plot distribution in $(X,Y)$ space. 
The left-right asymmetric distribution of the Dalitz plot can be obtained through subtracting each of the binned data at the $X>0$ side by the other side with opposite $X$ and the same $Y$, i.e.
\begin{equation}
    N_i^{\not C}(X_i, Y_i) = \frac{1}{2}[N_i(X_i, Y_i) - N_i(-X_i, Y_i)], ~~~~(X_i>0).\label{eq:N_asy}
\end{equation}
where $(X_i, Y_i)$ is the coordinate of the center of the $i$-th bin 
and $N_i(X_i, Y_i)$ represents the number of events in 
bin $i$.
The asymmetric events 
$N^{\not C}_i(X_i, Y_i)$ can be related to the theoretical amplitudes as
\begin{equation}
    \frac{N^{\not C}_i}{N_{\text{tot}}}=\frac{\int_i2\text{Re}\left[\mathcal{M}^C(X,Y)\cdot\mathcal{M}^{\not C}(X,Y)^\ast \right]d X d Y}{\int |\mathcal{M}^C(X,Y)|^2dXdY  },\label{eq:fit_binned}
\end{equation}
where $N_{\text{tot}}$ represents the total number of events in the whole phase space and $\int_i$ means that the integral region is within the $i$-th bin.
For the denominator on the right-hand side, since the integral region is the whole phase space, only the $C$-conserving part of the amplitude square survives, and we apply the NLO ChPT form~\cite{Gasser:1984pr} as done in Ref.~\cite{Gardner:2019nid} in our fit.

\begin{figure}[htbp]
    \centering
    \includegraphics[width = 0.42\textwidth]{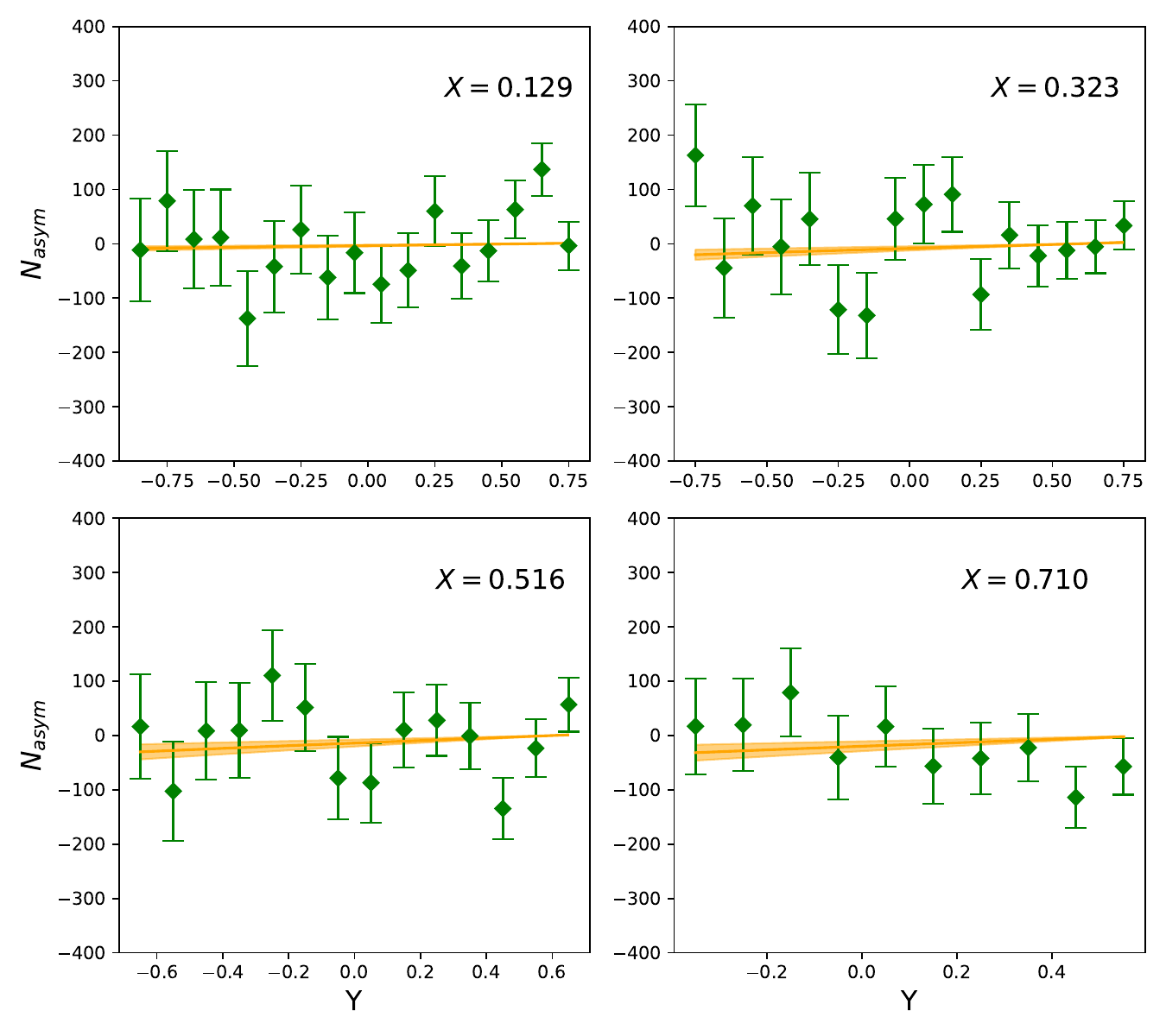}
    \hspace{0.8cm}
    \includegraphics[width = 0.42\textwidth]{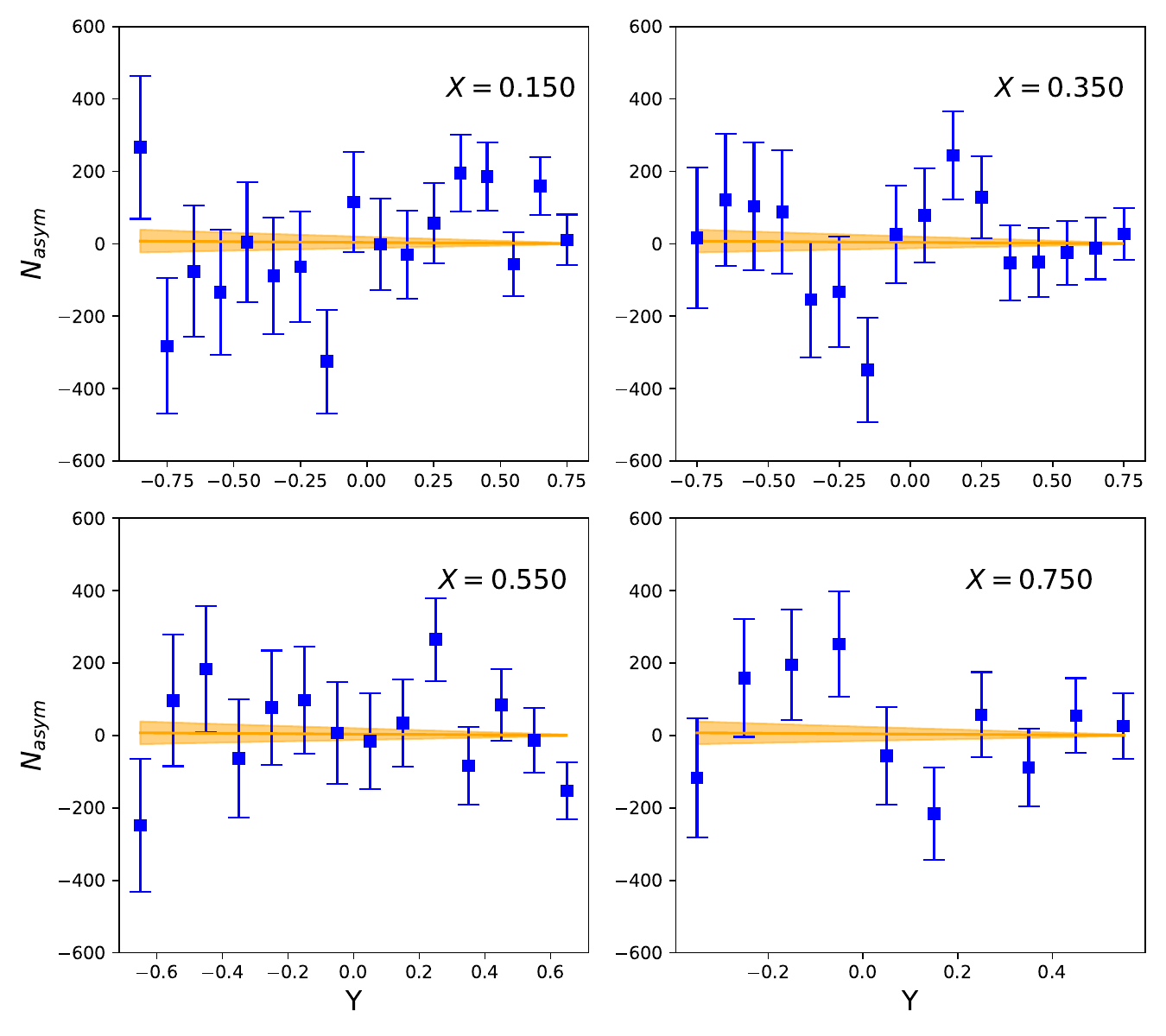}
    \caption{The comparison of the experimental left-right asymmetric distribution with the theoretical result
    using the order $p^2$ amplitude in Eq.~(\ref{eq:amp_p2}), with
        $\alpha$ determined from fitting the KLOE-2~\cite{KLOE-2:2016zfv} (left panel) and the BESIII~\cite{BESIII:2023edk} (right panel) data separately.
    }
    \label{fig:fit_bin_p2}
\end{figure}

We use Eq.~(\ref{eq:fit_binned}) to fit the KLOE-2~\cite{KLOE-2:2016zfv} and BESIII~\cite{BESIII:2023edk} data.
For $\mathcal{M}^{\not C}$ in Eq.~(\ref{eq:fit_binned}), we either use the order $p^2$ $I=2$ amplitude only or the order $p^2$ amplitude together with the order $p^6$ $I=0$ amplitude.
Adding the order $p^6$ $I=0$ amplitude to the order $p^2$ amplitude is in principle not proper
since the order $p^4$ amplitude has 
not been included.
However the order $p^6$ amplitude is a higher-order correction, and we include it to see roughly what the $I=0$ effects might be.
Figure~\ref{fig:fit_bin_p2} illustrates the fitting with $p^2$ amplitude only for 
different 
selected $X$.
The data points are $N_i^{\not C}$ obtained from the experimental data using Eq.~(\ref{eq:N_asy}), and the yellow line represents the fitted theoretical $C$-asymmetric distribution with the band indicating the uncertainty at $\pm 1\sigma$.
One can see that both the KLOE-2 data (left panel) and the BES-III (right-panel) data are basically consistent with zero and that the BES-III data have larger errors, which is understandable since they involve smaller $\eta$ decay 
 samples.
The fitted lines are also consistent with zero within two sigma.
All the fitting results together with the $\chi^2/\text{d.o.f.}$ are listed in Table~\ref{tab:alpha_beta_AQ_AS}.
Similarly to what we have learned from Fig.~\ref{fig:fit_bin_p2}, the values of $\alpha$ and $\beta$ are consistent with zero within two sigma.
In this case, their uncertainties are more significant than the central values and can be treated as the upper limits of the experimental constraint.
Again, the errors of $\alpha$ and $\beta$ from the BESIII data are about 3 times of those from KLOE-2, which is as expected since the statistics of BESIII is about $1/10$ of KLOE-2.
The results from BESIII and KLOE-2 are consistent within errors, so the joint fitting is feasible.
One can also conclude that adding the order $p^6$ amplitude has little effect on the $\alpha$ values, which means the order $p^6$ amplitude does indeed 
behave as a 
higher order 
effect. 
The large uncertainties in $\beta$ mean that 
the experimental data provide much less constraint on the decay 
to the $I=0$ final state.

\begin{table}[h]
\caption{Determined values of $\alpha$ and $\beta$ together with the $\chi^2/\text{d.o.f.}$ by fitting the left-right asymmetric distribution of KLOE-2~\cite{KLOE-2:2016zfv} and BESIII~\cite{BESIII:2023edk}.
}\label{tab:fit1}
    \begin{tabular}{c|c|c|c}
    \hline
~ & KLOE-2 & BESIII & KLOE-2 + BESIII \\
     \hline
     $\alpha/\text{GeV}^{-2}$ & $-0.031(14)$ &  $0.009(36)$ & $-0.026(13)$\\
         $\chi^2/\text{d.o.f}$ & $1.00$ & $0.97$ & $0.99$\\
\hline
$\alpha/\text{GeV}^{-2}$ & $-0.033(15)$ & $0.011(37)$ & $-0.027(14)$\\
$\beta/\text{GeV}^{-6}$  & $-7(12)$ & $14(29)$ & $-4(11)$\\
        $\chi^2/\text{d.o.f}$ & $1.00$ & $0.97$ & $0.99$ \\
     \hline
\end{tabular}
\end{table}

\begin{figure}[h]
    \centering
    \includegraphics[width = 0.4\textwidth]{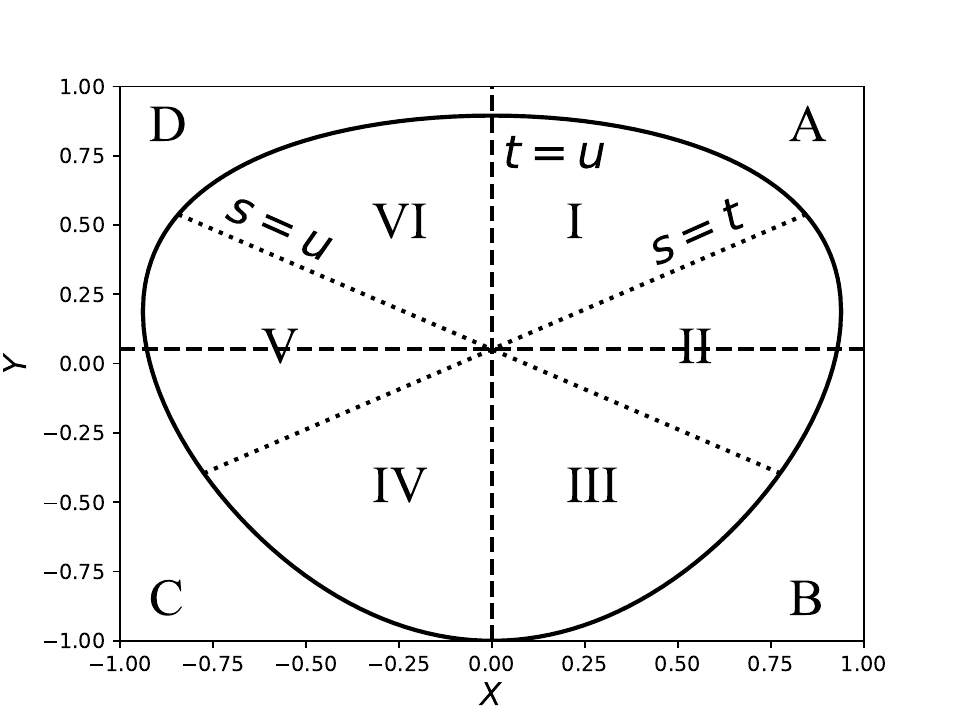}
    \caption{
    Different partitions of the Dalitz plot data
    used in the quadrant $A_{Q}$ and sextant $A_{S}$ asymmetries defined in Eqs.~(\ref{eq:A_Q_def}) and (\ref{eq:A_S_def}), respectively.\label{fig:Dp_section}}
\end{figure}

Now we switch to the second procedure in which we use
the integrated charge asymmetries $A_{Q}$ and $A_{S}$.
The quadrant asymmetry $A_Q$ and the sextant asymmetry $A_S$ of the $\eta\to\pi^+\pi^-\pi^0$ decay are defined as
\begin{eqnarray}
    A_Q &=& \frac{N_A + N_C - N_B - N_D}{N_A + N_C + N_B + N_D},\label{eq:A_Q_def}\\
    A_S &=& \frac{N_I + N_{III} + N_{V} - N_{II} - N_{IV} - N_{VI}}{N_I + N_{III} + N_{V} + N_{II} + N_{IV} + N_{VI}}\label{eq:A_S_def},
\end{eqnarray}
where the different partition
are illustrated in Fig.~\ref{fig:Dp_section}.
we can use $A_Q$ to deduce $\alpha$ and $A_S$ to determine $\beta$,  respectively.
KLOE-2~\cite{KLOE-2:2016zfv} and BESIII ~\cite{BESIII:2023edk} report their measurements of $A_{Q}$ and $A_{S}$ as: $A_{Q}^{\text{KLOE}} = (+1.8\pm 4.5^{+4.8}_{-2.3})\cdot 10^{-4}$, $A_S^{\text{KLOE}}=(-0.4\pm 4.5^{+3.1}_{-3.5})\cdot 10^{-4}$, $A_{Q}^{\text{BES}} = (-3.5\pm 13.1\pm1.1)\cdot 10^{-4}$ and $A_S^{\text{BES}}=(-7.0\pm 13.1\pm 0.9)\cdot 10^{-4}$.
Accordingly, we use the expression
\begin{equation}
    N_X = \int_X |\mathcal{M}(X,Y)|^2dXdY,
\end{equation}
where $X$ denotes the partitions, and Eq.~(\ref{eq:A_Q_def}) and (\ref{eq:A_S_def}) are used to calculate the theoretical asymmetries with $\alpha$ and $\beta$ to be determined. 
The resultant $\alpha$ and $\beta$
are shown in Table~\ref{tab:alpha_beta_AQ_AS}.
One can see that the $\alpha$ values are consistent with those from fitting the left-right asymmetric distribution within errors, 
though 
the uncertainties are slightly larger in this procedure.
Determining $\beta$ from $A_S$ can, in principle,  result in a better signal since only the order $p^6$ amplitude is included. However, as we see,  the resultant value is not 
better constrained. 
This again suggests that the present data cannot yet provide a precise 
constraint on the $I=0$ channel.

\begin{table}[h]
\caption{Values of $\alpha$ and $\beta$ determined by $A_S$ and $A_Q$ from KLOE-2 and BESIII, respectively.
\label{tab:alpha_beta_AQ_AS}}
    \begin{tabular}{c|c|c}
    \hline
~ & KLOE-2 & BESIII \\
     \hline
     \rule{0pt}{1em}
  $\alpha/\text{GeV}^{-2}$ ($A_Q$) & $-0.007^{+0.019}_{-0.023}$ & $0.013\pm 0.048$\\
  \hline\rule{0pt}{1em}
  $\beta/\text{GeV}^{-6}$ ($A_S$) & $-1.05^{+14.93}_{-14.61}$ & $18.33\pm 33.66$ \\
  \hline
\end{tabular}
\end{table}

We have applied two procedures to determine $\alpha$ and $\beta$. 
We use the values determined by jointly fitting the KLOE-2 and BESIII left-right asymmetric distribution as our primary results, i.e., $\alpha=-0.027(14)~\text{GeV}^{-2}$ and $\beta=-4(11)~\text{GeV}^{-6}$.
For the BSM scale estimation in the next section,
we estimate the upper limits of their absolute vales at $90\%$ confidence level (C.L.) as
\begin{eqnarray}
    |\alpha| &\lesssim& 0.05 ~\text{GeV}^{-2},\\
    |\beta| &\lesssim & 22~\text{GeV}^{-6}.\label{eq:value_ab}
\end{eqnarray}

Given the limitations of our 
current data 
in determining a constraint
on $I=0$ 
BSM effects, we pause to consider how else these 
effects could be
constrained. Here we note
the possibility of 
constraining the $I=0$ sector 
through the study of 
$\eta\to \pi^0 \ell^+ \ell^-$ decay, 
though the current experimental limits on 
the
latter give 
comparable 
constraints~\cite{Akdag:2023pwx}. 

\section{New Physics Scale and Impact Study}\label{sec:new_physics_scale}

\begin{figure}[htbp]
    \centering
    \includegraphics[width = 0.6\textwidth]{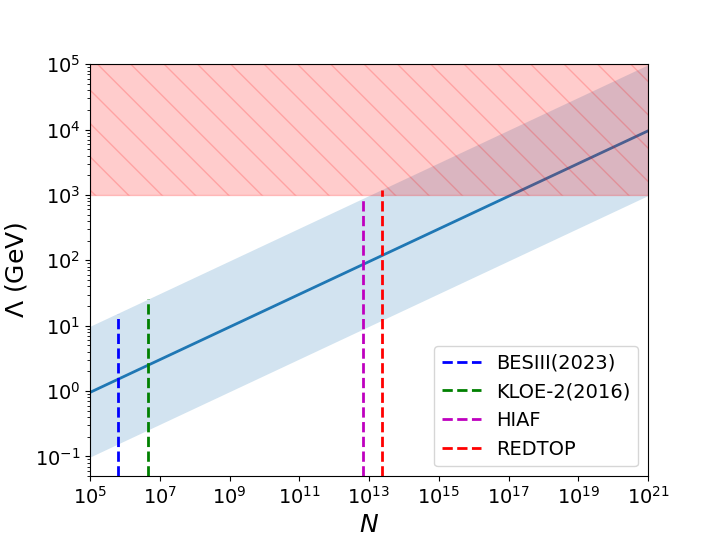}
    \caption{The predicted relation between estimates 
    of new physics scale and experimental statistics 
    for $\eta\to\pi^+\pi^-\pi^0$ decay, 
    where the pink area with slashes means the energy region where new physics could occur, estimated to be $\Lambda>1$~TeV, and the blue band is our theoretical result with uncertainty. The blue, green, purple, and red dashed lines indicate the statistics from the BESIII~\cite{BESIII:2023edk}, KLOE-2~\cite{KLOE-2:2016zfv}, HIAF~\cite{Chen:2024wad},  and REDTOP~\cite{Gatto:2016rae,Gatto:2019dhj,REDTOP:2022slw} experiments, respectively.}\label{fig:N_Lambda}
\end{figure}

With the upper limits of $|\alpha|$ and $|\beta|$, we utilize NDA for a rough order-of-magnitude estimate 
of the new physics scale.
When matching the dimension 7 quark-level operator in Eq.~(\ref{eq:quark}) to meson-level ones, an effective low energy constant 
multiplies each operator, and according to NDA~\cite{Manohar:1983md, Weinberg:1989dx, Georgi:1992dw, Jenkins:2013sda, Gavela:2016bzc,Manohar:1996cq} its natural size is expected to be  
\begin{equation}
    F_0^4\Lambda_\chi^3\frac{1}{F_0^m}\frac{1}{\Lambda_\chi^n},
\end{equation} 
where $m$ is the number of meson fields and $n$ denotes the number of derivatives in the ChPT operator.
Thus for the order $p^2$ operators in Eq.~(\ref{eq:p2_ChPT}), the coefficient $g_i$ has the order of ${F_0^4}{\Lambda_\chi}$.
Analogously, for the operators at order $p^6$ in Eq.~(\ref{eq:ChPT_p6}), $f_i\sim {F_0^4}/{\Lambda_\chi^3}$.
Assuming there is no unexpected fine tuning in the ultraviolet completion of SMEFT, $c_{ij}$ should have an order of unity.
Therefore, from Eq.~(\ref{eq:Neta_p2}) and Eq.~(\ref{eq:Neta_I0}), we have $\mathcal{N}_{p^2}\sim {F_0^4}{\Lambda_\chi}$ and $\mathcal{N}_{p^6}\sim {F_0^4}/{\Lambda_\chi^3}$.
Based on Eq.~(\ref{eq:amp_p2}) and Eq.~(\ref{eq:amp_p6}), the new physics scale is finally related to $\alpha$ and $\beta$ as
\begin{eqnarray}
 \Lambda_{p^2} &\sim & {\left(\frac{vg\Lambda_\chi}{ M_W^2}\frac{1}{|\alpha|} \right)}^{1/2},\label{eq:L_alpha}\\
    \Lambda_{p^6} &\sim & {\left(\frac{vg}{ M_W^2\Lambda^3_\chi}\frac{1}{|\beta|} \right)}^{1/2},
    \label{eq:L_beta}
\end{eqnarray}
where $\Lambda_{p^2}$ and $\Lambda_{p^6}$ represents the new physics scale derived from the coefficients of $p^2$ and $p^6$ amplitudes, respectively.
Note that our resulting 
$\Lambda$ is proportional to $-1/2$ power of the amplitude coefficients which is different from the LEFT work of Ref.~\cite{Akdag:2022sbn}.
In Sec.~\ref{sec:method} we estimate the upper limits of $|\alpha|$ and $|\beta|$ at $90\%$ C.L. as $|\alpha|\lesssim 0.05\,\rm GeV^{-2}$ and $|\beta|\lesssim 22\, \rm GeV^{-6}$.
Utilizing Eq.~(\ref{eq:L_alpha}) and Eq.~(\ref{eq:L_beta}) we have
\begin{eqnarray}
 \Lambda_{p^2} & \gtrsim & 0.8 ~ \text{GeV},\\
    \Lambda_{p^6} & \gtrsim & 0.03 ~\text{GeV},
    \label{eq:value_L_beta}
\end{eqnarray}
where we use $v=246.22$ GeV, $M_W=80.37$ GeV, $g=0.653$,  and $F_0=92.28$ MeV as 
reported in 
Ref.~\cite{ParticleDataGroup:2022pth}, noting that their empirical errors are unneeded 
since we care only about the order of magnitude. 
It is understandable that $\Lambda_{p^6}$ is very small, because as mentioned in Sec.~\ref{sec:method}, the current experimental data have little constraint on $\beta$.
The scale
of 
the $p^2$ coefficient is much larger than that
of the $p^6$ coefficient, but still obviously lower than the expected new physics scale, which implies that the current experiments 
are not yet precise enough to provide a reasonable 
estimate 
of the new physics scale.
However, our analysis is valid to estimate the possibility that the $C$ and $CP$ violation of $\eta\to\pi^+\pi^-\pi^0$ could be observed by the upcoming $\eta$ experiments.
An impact study from the order $p^2$ coefficient is performed as follows, from which one can see that the low limits of $\Lambda$ are driven by the relevant low statistics of present data as shown in Fig.~\ref{fig:N_Lambda}.

On account that the value of $\alpha$ is statistically consistent with zero and the uncertainty indicate the experimental precision, we use its upper limit using its uncertainty as $|\alpha|\lesssim 0.1~\text{GeV}^{-2}$ in the impact study.
Thus the current estimation of new physics scale would be $\Lambda \gtrsim 1.7 ~ \text{GeV}$.
Since the scale is just a rough order-of-magnitude approximation from 
NDA, it is reasonable to consider
a $\pm 1$ order of magnitude deviation as 
its 
uncertainty.
Suppose, e.g.,  $\Lambda_{p^2}$ 
should increase from $10$ GeV to $1$ TeV, then, according to Eq.~(\ref{eq:L_alpha}), the upper limit of $\alpha$, which is estimated by its uncertainty, should decrease 
by a factor of 
$10^{-4}$. 
Since the uncertainty of $\alpha$ is proportional to
the experimental statistics 
should improve by $10^8$ --- at least.
Generally, if we think 
$\Lambda$ 
should 
increase by $10^{n}$ times, the experimental statistics would need to improve by $10^{4n}$ times.
Note that, as 
pointed out 
in the first section, for a pure $CP$-violating process,
rather than the interference
effect we study in 
$\eta\to\pi^+\pi^-\pi^0$ decay, 
the required experimental improvement would be $10^{8n}$ times.

The data we use in this analysis are from the KLOE-2 collaboration~\cite{KLOE-2:2016zfv} with a sample of $4.7\times 10^{6}$  $\eta\to\pi^+\pi^-\pi^0$ decays and BESIII collaboration~\cite{BESIII:2023edk} with a sample size of $6.3\times 10^{5}$ events. 
Supposing that the new physics
lives at a scale above 1~TeV,
it means future experiments need to reach a sample size of at least $10^{13}$ to observe $C$ and $CP$ violation in $\eta\to\pi^+\pi^-\pi^0$ decay.
We illustrate the relation between experimental statistics and the estimated new physics scale in Fig.~\ref{fig:N_Lambda}, in which the pink area with slashes means the energy region where new physics could occur and the blue band is our theoretical result with uncertainty.
Future experiments are expected to be made at the JLab eta factory(JEF) of Jefferson lab~\cite{Gan:2009zzd,Gan:2017kfr,Somov:2024jiy}, by the REDTOP collaboration~\cite{Gatto:2016rae,Gatto:2019dhj,REDTOP:2022slw} 
and by the collaboration from the HIAF eta factory~\cite{Chen:2024wad}, etc.
Among them, the REDTOP collaboration plans to have 
as much as $10^{14}$ $\eta$ events in three years of running, or about $2.3\times 10^{13}$ $\eta\to\pi^+\pi^-\pi^0$ samples.
The HIAF collaboration~\cite{Chen:2024wad} plans to have over $10^{13}$ $\eta$
events 
per year, which is about $6.9\times 10^{12}$ $\eta\to\pi^+\pi^-\pi^0$ events 
for a three-year period of running. 
We draw the statistics of BESIII~\cite{BESIII:2023edk}, KLOE-2~\cite{KLOE-2:2016zfv}, HIAF~\cite{Chen:2024wad}, and REDTOP~\cite{Gatto:2016rae,Gatto:2019dhj,REDTOP:2022slw} 
from left to right 
in Fig.~\ref{fig:N_Lambda}, respectively.
One can see that the 
new-physics 
potential of studies of $C$ and $CP$ violation is promising and could possibly 
be observed in the upcoming HIAF and REDTOP experiments.

\section{Summary and Outlook} \label{sec:summary}
The decay $\eta\to\pi^+\pi^-\pi^0$ is an ideal process 
in which to study 
flavor-conserving $C$ and $CP$ violating physics BSM. 
We have shown the key procedures involved in deriving the quark-level operators pertinent to $C$- and $CP$-violation in $\eta\to\pi^+\pi^-\pi^0$ decay, originating from the dimension 6 SMEFT operators.
Then,
besides the order $p^2$ meson-level $C$- and $CP$-odd operators corresponding exclusively to the $I=2$ final state,
the order $p^6$ operators producing non-vanishing $I=0$ amplitude at lowest order are presented for the first time.
By directly fitting the asymmetric distribution of the Dalitz plot of $\eta\to\pi^+\pi^-\pi^0$ from KLOE-2~\cite{KLOE-2:2016zfv} and BESIII~\cite{BESIII:2023edk} collaborations,
the upper limits of the order $p^2$ and $p^6$ amplitude coefficients $\alpha$ and $\beta$ from using 
chiral effective theory are determined without 
theoretical model 
uncertainties. 
Based on NDA, we perform an order of magnitude estimate
of the new physics scale.
A corresponding impact study of future $\eta$ experiments  is also carried out and shown in Fig.~\ref{fig:N_Lambda}.
Considering future experiments with much higher statistics, such as planned by the REDTOP collaboration~\cite{Gatto:2016rae,Gatto:2019dhj,REDTOP:2022slw} and at the HIAF eta factory~\cite{Chen:2024wad}, it seems
promising that 
$C$- and $CP$-violation in $\eta\to\pi^+\pi^-\pi^0$ decay could be observed in the near future.

\begin{acknowledgments}
We are grateful to Xiaodong Ma, Yi Liao, Hakan Akdag, Tobias Isken, Bastian Kubis, Shuang-Shi Fang, Xiao-Lin Kang, Xu-Rong Chen and Qian Wang for helpful discussions.
This work is partially supported by Guangdong Major Project of Basic and Applied Basic Research under Grant No.\ 2020B0301030008.
JS is supported by
the Natural Science Foundation of China under Grant No. 12105108.
JL is supported by the Natural Science Foundation of China under Grant No.\ 12175073 and No.\ 12222503, and the Natural Science Foundation of Basic and Applied Basic Research of Guangdong Province under Grant No.\ 2023A1515012712.
SG acknowledges partial support from the U.S. Department of Energy, Office of Science, Office of Nuclear Physics 
under contract DE-FG02-96ER40989.
\end{acknowledgments}

\bibliography{library}

\end{document}